\documentclass{article}
\usepackage[utf8]{inputenc}
\usepackage{authblk}
\usepackage{setspace}
\usepackage[margin=1.25in]{geometry}
\usepackage{graphicx}
\graphicspath{ {./figures/} }
\usepackage{subcaption}
\usepackage{amsmath}
\usepackage{lineno}
\usepackage{booktabs, multirow} 
\usepackage{changepage,threeparttable} 

\usepackage{CJK}
				\newcommand{\zh}[1]{\begin{CJK}{UTF8}{bsmi}#1\end{CJK}}

\usepackage[style=nejm, 
citestyle=numeric-comp,
sorting=none]{biblatex}
\title{Identity Collapse? Realignment of Taiwanese Voters in the 2024 Presidential Elections on Social Media}

\author[1*]{Ho-Chun Herbert Chang}
\author[2]{Sunny Fang}

\affil[1]{Department of Quantitative Social Science, Dartmouth, Hanover NH}
\affil[2]{Department of Computer Science, Barnard College, New York NY}
\affil[*]{Address correspondence to: herbert@dartmouth.edu}

\date{}

\begin{document}

\maketitle

\begin{abstract}
The 2024 Taiwanese Presidential Election is not just a critical geopolitical event, it also engages with long-standing debate in politics regarding the factors that lead to the rise of new political parties and candidates. In 2021, the Economist called Taiwan ”the most dangerous place on earth” due to its critical role in a fragile supply chain. Additionally, a four-candidate race has emerged in a traditionally bipartisan election which begs the question: how will voters realign given the choice of four candidates? Leveraging more than a million posts on social media, we analyze user (predominantly Taiwanese) discourse and engagement along the axes of national identity, issue topic, and partisan alignment. Results reveal alternative candidates (Ko and Gou) draw attention from the fringes rather than the center relative to national identity, and traditional candidates derive more engagement from the traditional media and salience to geopolitical issues. Crucially, in-group references generate more engagement than out-group references, contrary to Western-based studies. We discuss how the dissolution of Taiwan’s single-issue society may not just lead to more viable candidates and multi-issue discourse, but the misalignment of national and partisan identity may heal deep-seated partisan cleavages.


\end{abstract}


\section{Introduction}

Dubbed "the world's most dangerous place" by the Economist in 2021~\cite{Economist_2021,rigger2021people}, Taiwan has become a geopolitical lynchpin due to its contentious history with China and its role in producing the world's Silicon Chips~\cite{miller2022chip,hollihan2023public}. As a result, the 2024 Taiwanese Presidential Elections have not just captured the attention of local voters, but significant interest from countries around the world as the foreign policy enacted through Taiwan's presidency will have significant ramifications given heightened aggression from China. For political scientists, this election also presents an interesting case-study of voter realignment~\cite{shafer1991end,abramowitz1998ideological} especially in the context of growing polarization in the United States and Europe~\cite{stonecash2018diverging,gidron2022many}. Additionally, for the first time, a four-person race has emerged in a traditionally bipartisan Presidential Electorate. It allows us to ask a simple question: how does voter attention shift when the number of candidates doubles?

In past elections, the pan-Blue alliance led by the Kuomintang (KMT) and pan-Green alliance led by the Democratic People's Party (DPP) embodied competing visions of the nation's identity--- Taiwan versus the Republic of China (R.O.C.). However, in the aftermath of the Hong Kong protests~\cite{rigger2022unification}, recent polls claim that there is a shift away from the "China Problem" to core domestic issues, such as labor law, housing affordability~\cite{Cheng_2023}, the rights of migrant workers, and sexual harassment in the work place. Political scientists have theorized the maximum number of political parties depends on the number of societal cleavages~\cite{cox1997making} even in mixed electoral systems~\cite{cox1994seat}; the Taiwanese 2024 Presidential election serves as an infrequent opportunity to study this phenomenon. At a time when the issue space is diverse yet malleable, how voters realign to new parties and new candidates is of crucial interest. 

In this paper, we investigate the supply-demand dynamics of the four presidential candidates of Taiwan, using 911,510 Facebook posts from public figures, pages, and public groups, provided by CrowdTangle. We present three key findings. 
First, we find the two traditional candidates Lai and Hou generate more engagement from the traditional media (with Hou surprisingly the greatest coverage), converge on geopolitical issues more, and are central in terms of national identity. In other words, alternative candidates come from the fringe, which shows for this election independents may be moderate~\cite{wang2019myth} in regard to partisan identity but not national identity. Second and crucially, in-group mentions drive greater engagement rather than out-group comparisons, contrary to American congressional dynamics~\cite{rathje2021out}. However, results align where in-group references predict positive affect whereas out-group references produce negative affect. Third, and perhaps most importantly, attitudes toward China do not predict strong virality. Rather, relations with the United States and technology are the key discourse topics, which indicates a shift in how the Taiwanese perceive themselves within geopolitical debates.

\subsection{A brief history of Taiwanese National Identity}
National identity, commonly bound by the ethnic and civic commonalities shared by a nation, is also shaped by the existence of a “significant other,” another entity with geographical proximity and perceived to be a threat to the nation's independence~\cite{triandafyllidou1998national}. Given its history, Taiwan's national identity has been dominated by "the China factor": citizens constantly reconcile their undeniable cultural inheritance from China and distinguished their Taiwanese identity by expressing their opposition to China’s authoritarian government, to construct a national "in-group" and "out-group."

After the first Sino-Japanese War in 1895, Taiwan came under Japanese colonization until the end of World War II. In the aftermath, Japan ceded its control of Taiwan to the Republic of China, then ruled by the Chinese Nationalist Party--also known as Kuomintang (KMT). Under KMT's rule, an ethnic cleavage emerged between the "mainlanders" (\textit{waishengren}) and "Taiwanese" (\textit{benshengren}), which was deepened as a result the 1947 “228 Incident”--an anti-government uprising that ended in a KMT-led massacre~\cite{corcuff2016taiwan}. In 1949, Chiang Kai-shek, the then-KMT leader, retreated to Taiwan after their defeat by the Chinese Communist Party (CCP). Initially treating Taiwan as a mere military base and in hopes of reclaiming the mainland someday, the Chiang regime attempted to “re-Sinicize” residents of the island. Emphasizing cultural, ethnic, and historical ties to China, the KMT induced a Chinese identity--albeit more saturated--among the Taiwanese population that lives on until today. 

In the 1980s to 1990s, Taiwan underwent a period of rapid economic growth and democratization. When Lee Teng-hui became the first native-born to assume office in 1988, he lifted the martial laws and legalized political parties. These constitutional reforms, along with his other efforts, diluted the mainland legacy crafted by the Chiang regime and cultivated a more ethnically inclusive identity~\cite{wang2004contending}. Despite sociopolitical advancements, a sense of discordance emerged among citizens due to the lack of recognition in the international sphere~\cite{achen2017taiwan}. The interplay of democratization at home and ostracization abroad called for the need to utilize public diplomacy to enhance Taiwan's international recognition as a progressive liberal state, in contrast to China's authoritarianism. These series of development led to the engendering of the Taiwanese identity. 

Taiwanese identity was not the only product of Taiwan's democratization; what also ensued were multiple political parties derived from the KMT, replacing the single-party system with multiparty one. In 2000, Chen Shui-bian from the Democratic Progressive Party (DPP) won the presidential seat, serving as the impetus of the emergence of the Pan-Blue Alliance and the Pan-Green Alliance, with KMT and DPP in the lead, respectively~\cite{achen2017taiwan}. The co-occurrence of these two events produced an associative relationship between the DPP and the Taiwanese identity. As a growing population began to renunciate their Chinese identity, these Taiwanese-identifying people leaned towards the cause of Taiwanese independence and the DPP, which espoused the idea in the 2000s~\cite{liu1999taiwanese}.

The cleavage can thus be characterized as such: Pan-Blue indicates allegiance to R.O.C., also known as the Chinese consciousness. Pan-Green indicates allegiance to Taiwan as national identity, or the Taiwanese consciousness. Unlike the USA where party alignment is with conservative versus liberals~\cite{levendusky2009partisan}, national identity is the focal point of political cleavage. During Chiang's regime, ethnic divides between the mainlanders and the local citizens drove polarization. After democratization, partisan attachment is rooted in their identification of either Chinese or Taiwanese consciousness. Today, with China continuing posing as both a security threat and cultural heritage, "the China factor" remains a decisive pivot that shapes the voter's choice. 

As demonstrated with Taiwan's democratic development, traditionally there exists a strong association between party identification and national identity. Through a constructionist lens, Taiwan's national identity can be defined with artificial boundaries and crafted via deliberate efforts~\cite{wang2017changing}. The emergence of two additional candidates in Ko and Gou then begs the question of not just how these independent parties emerged (such as the TPP), but also their national alignment~\cite{Smith_2023}. Just as how emergent independents in the 1990s embodied unique and diverse interpretations of national identity, independents caught in shifting geopolitical tides today face the task of doing the same.


\subsection{The Missing Middle? The Taiwanese Voter}
In the study of democratic voting, three big themes are thought to factor into citizen decision making---party identification, issue orientation, and candidate evaluation~\cite{jacoby2010policy}. However, as Achen and Wang contend, the case of Taiwan should include a fourth category: national identity.
In 2017, the "China factor" was the most important political division in the country's politics. Then, the cleavage between pan-Blue and pan-Green voters concerned the relationship with China, so much that Taiwan was often regarded as a single-issue along a central axis of unification and independence. However, even before the protests in Hong Kong, the orientation of voters had begun to shift. According to annual surveys administered by the Election Studies Center in Taiwan, more than half of Taiwanese citizens considered themselves Taiwanese by 2009 (51.6\%) which had risen to 62.3\% by 2021~\cite{TaiwaneseChineseIdentity2023}. A greater cultivation of Taiwanese identity is attributed to the youth who only grew up in a democratic Taiwan and the Sunflower Movement.

The "freezing hypothesis"--developed by Lipset and Rokkan~\cite{linz1967party}--noted that European party systems stabilized when they reaffirmed preexisting cleavages. While cleavages are important for democracies (Lipset contends moderate tension among contending political forces is necessary), societies that allow these cleavages to deepen also risk irredeemable polarization. The case of Taiwan with its four new candidates provides a case-study for how partisan divide can be "thawed." Notably, this is not the first time that parties outside of the big two have run for presidential election. In the 2000 presidential election, the KMT suffered an internal split for the first time, with contenders who were previously part of the Pan-Blue Alliance--such as James Soong--running as an independent~\cite{yu2017parties}.   

Wang highlights the importance of a growing nonpartisan population since 2015, citing results from the Taiwan National Security Survey (TNSS) ($n=11,660$). He observed that voters who are slightly pro-independence or pro-unification no longer pledge allegiance for either of the two dominant parties~\cite{wang2019myth}. The emergence of an independent moderate is crucial for improving democracy. It signifies that partisan appeals, namely on the emphasis on national identity (independence/unification), may not be as effective as in the past. Once the polarization on national identity has been relaxed, one could better understand what is at stake on the domestic issues landscape. 

\subsection{Domestic policy issues in Taiwan}
According to data from the Election Study Center at National Chengchi University (NCCU), eight items on the issue agenda were considered from the voters' perspective: economic prosperity, cross-strait affairs, wealth distribution, political corruption, national security, social reform/stability, and environmental protection. 
Four of the eight issues---wealth distribution, environmental protection/economy, reform/stability, and independence/unification---were cited as the focal issues by Sheng and Liao~\cite{sheng2017issues}. In the study of elections from 1996 to 2008, it was found that voters conscious of social welfare issues more likely voted for the DPP ~\cite{sheng2013issues}. However, wealth distribution, overshadowed by political affiliations, never grew into a significant cleavage. Similarly, economic prosperity usually took precedence over environmental protection in the public space, and the same trend is observed in both dominant parties. For issues where the views of partisan elites usually aligned with those of the voters, significant political cleavage was unlikely. On the two other issues, parties that emerged toward the end of the authoritarian era, most notably the DPP, emphasized policy reforms and shifted focus to the independence question. Alternative parties are instead characterized by focus on domesitic issues and routinization~\cite{nachman2023routine}.

The independence/unification question has been the most salient issue that drives political cleavage, serving as a position issue more likely to shape party competition~\cite{stokes1963spatial}; it is also a latent confounder and has the power to polarize issues. For instance, voters' support for the Economic Cooperation Framework Agreement (ECFA), a landmark trade deal between Taipei and Beijing, hugely depended on their identification with either the Pan-Green and Pan-Blue Alliance, with the working class being more likely to oppose and business elites tending to support the accord~\cite{lin2011cross}. If the trend persists to develop, the cleavage between independence/unification could fuel polarization among different classes.  

The ironic and cynical interpretation is that Taiwanese interest in domestic issues may stem from a need to differentiate parties, given a gradual collapse of a single-issue society and alignment in attitudes toward China. In other words, party preferences, reinforced by one's national identity, played a vital role in previous election cycles~\cite{lee2017we}. Between 2014 and 2018, there was a decline in "Taiwanese" identity. Wang attributes the decline to \textit{issue ownership + hedging}, explaining identity shifts based on the incumbent's performance~\cite{wang2023does}. If a pro-independence government performs poorly, then Taiwanese identity decreases; similarly, if a pro-unification government performs well, Taiwanese identity also decreases. As previously mentioned, another key factor is a "missing middle" in Taiwanese politics--a group of moderate independents that do not strongly identify with either side of national identity. These two observations help us understand why domestic issues are also at stake this election. Other than the eight issues noted in the NCCU survey, domestic issues such as the recent \#MeToo movement may matter, since the first victim who spoke out belonged to the incumbent party, the DPP~\cite{Wei_2023}.

Aside from national surveys, media outlets also show what issues garner the most attention. In analyzing Taiwan's three largest national newspaper groups (Liberty Times, China Times, and Apple Daily) and the most popular online-only outlet (ET Today), Rauchfleisch et al. (2022) found that China and social media ranks 1, followed by economic policy, and Taiwanese sovereignty~\cite{rauchfleisch2022taiwan}, which echoes how issues related to national identity take precedence over domestic issues.

\subsection{Group Dynamics on Social Media}

In the United States, 80\% of Americans receive some news and information from social media and this parallel growth occurs in most developed countries~\cite{newman2023digital,ortiz2023rise}. Although the rise of information technologies is  ubiquitous, there are meaningful differences in a country's digital information environment based on the dominant social media platforms, as the design of specific social media platforms have downstream effects. Platforms like Twitter are open by nature. Anyone can follow anyone, and as a result the  structure is more akin to ”broadcasting~\cite{goel2016structural}." Instagram in contrast has a more localized structure--- many accounts require follows to see each other's content, and due to its visual nature is embedded in physical spaces~\cite{chang2022justiceforgeorgefloyd}. Twitter datasets often only report 3\% geolocated tweets, whereas on Instagram this is as high as 13\%. Finally, chatrooms such as WhatsApp and Line are the most private, requiring invitations or links to join. Overall, 89\% of Taiwanese users use Facebook, 85\% use Line, whereas a meager 27\% use Twitter~\cite{chang2021digital}. Line, in particular, features closed chatrooms outside the public eye. This places Taiwan in a very different online information environment than Western countries. 

A significant stream of research considers how social media contributes to polarization~\cite{huszar2022algorithmic,chang2023liberals,nyhan2023like,guess2023reshares}. Scholars are particularly interested in growing affective polarization, where disagreement isn't over issue topics but due to dislike of opposing parties~\cite{kingzette2021affective,druckman2019we}. In the American context, Rathje et al. (2021) found out-group animosity generates the most virality~\cite{rathje2021out} using posts by congress members. 
In our study, we expand to other Facebook "surfaces" that facilitate communication. In particular, public groups allow wider discussion by every day users, and as such, are likely the best proxy for understanding public discourse, as scraping private groups and personal timelines on Facebook is rife with ethical questions of consent.

\subsection{Research Questions}

Our goal is to investigate how group dynamics, issue topics, and conception of national identity drive engagement on social media during this election. We focus specifically on the misalignment of in- and out-groups for partisan and national identity.
We have the following research questions:
\begin{itemize}
    \item \textbf{RQ1:} How do the four candidates split the dichotomy of Taiwan-R.O.C. national identity? We hypothesize alternative candidates will align on the axis of national identity, due to correlates with age. 
    \item \textbf{RQ2:} How does the traditional media generate engagement for the four candidates? Since Taiwanese national identity correlates with age, we hypothesize that the Taiwan candidate to have less cross-cutting engagement whereas the R.O.C candidate to have more.
    \item \textbf{RQ3:} What are differences in topic salience across the four candidates? We hypothesize that alternative candidates will have greater engagement on domestic issues as traditional candidates will be defined by geopolitics owing to their parties' history.
    \item \textbf{RQ4:} In generating virality, what issues and group statuses produce the most engagement? We hypothesize alternative candidates will generate more virality via domestic issues, due to traditional candidates' proclivity toward geopolitics. Traditional candidates will also generate stronger party in-group references by appealing to their partisan identity.
\end{itemize}


\section{Data and Methods}

\subsection{Data Collection}
Our analysis follows two streams of data---social media and websites of legacy news media. For social media, we searched Facebook and Instagram using CrowdTangle, directly searching for elections-related discourse and the candidates. There was significantly more content on Facebook than Instagram, which adheres to previous findings about the less political nature of Instagram~\cite{chang2022justiceforgeorgefloyd} and the greater usage of Facebook in Taiwan, relative to Instagram. As previously mentioned, public Facebook groups are likely the best proxy for understanding public discourse, without scraping private groups and personal timelines on Facebook.

Table~\ref{tab:crowdtangle-searches} shows the types of searches we conducted. A list of keywords is included in the Appendix. When scraping the official candidate and party page, we discovered the DPP had separate pages for their official spokesperson, their youth representatives, and their legislative updates. In comparison, the KMT and TPP had one official page. We then took particular care to scrape posts from public groups, which consisted of 63.9\% of all the total posts. This gives a sample of the discourse. We used dates staring from 01/12/2023 to capture all content exactly a year prior to when the elections are held.
\begin{table}[!htb]
\begin{tabular}{|p{2in}|p{4in}|}
\hline
\textbf{Data Source}                         & \textbf{Description}                                                                                                                                                               \\ \hline
General Elections& Searched on keywords pertaining to the presidential election (i.e. \zh{大選}).\\ \hline 
 Official Candidate and Party Pages&Searched on each candidate's official page. This included Tsai Ing-Wen and Han Kuo-Yu due to their large presence in the overall dataset. \\\hline \hline
CrowdTangle Candidate General& Search on keywords by candidate. This includes all mentions of the candidate including in oppositional groups.\\ \hline
CrowdTangle Candidate and Identity Community & Using the Candidate General dataset, we produced a manual subset of public candidate support groups, i.e. \zh{挺賴團} (Lai support group). Also includes groups that explicitly mention Taiwan and R.O.C., such as \zh{中華民國粉絲團} (Republic of China Fan Club).\\ \hline
\end{tabular} \caption{List of Searches on CrowdTangle} \label{tab:crowdtangle-searches}
\end{table}
A full list of the search terms and groups are included in the Appendix. Tsai and Han were included as they were prior presidential candidates; prior work has shown the incumbent and prior candidates garnering significant referrals for the in-group and out-group.

\subsection{Topic Modeling and Toxicity Labeling}

To gauge topic salience for our third question, we extracted keywords from datasets for General Elections and CrowdTangle Candidate and Identity Community. To do so, we utilized jieba~\cite{jieba}, a Python Chinese word segmentation module. Since jieba was primarily trained on Simplified Chinese, the chinese-converter~\cite{chinese_converter} library was used to convert the text data from Traditional Chinese to Simplified Chinese before conducting keyword extraction. 

We first tagged the two datasets by three main categories: political figures (e.g., the four candidates), geopolitics (e.g., the USA), and domestic issues, which we adhered to the policy issues outlined by Sheng and Liao~\cite{sheng2013issues} and added relevant topics such as technological advancement, seen as Taiwan is often mentioned in the discussion of the global semiconductor race~\cite{miller2022chip}. After filtering the data sources by policy issues and candidates, we employed the two keyword extraction methods: frequency counts and Latent Dirichlet Allocation (LDA) for topic modeling. LDA works by selecting the number of clusters that yielded the largest coherence value~\cite{blei2003latent}. To measure toxicity, we use the Google Perspective API frequently used in political science research to operationalize incivility~\cite{frimer2023incivility}.

\subsection{In-/Out-Group Labeling}
To answer our last research question, we are interested in the role partisan in-/out-group messages have on virality. The criteria for in-group used is a reference to one's own party, themselves, and a previous presidential candidate of their own party. For instance, for Lai this would include Lai, Tsai, and the DPP. For out-group labeling, this is the full set of candidates and parties minus their own party and prior candidate. For instance, Ko would have Tsai, Lai, Han, Gou, Hou, the DPP, and the KMT (all except himself and the TPP). The only caveat to this labeling strategy is for Hou and Gou, who are in the same party. They are considered oppositional. As political ads can be generalized as "attack," "promote," and "contrast, assuming a similar intention but with a different political communication modality, this labeling process allows us to contrast these three intentions.

\section{Results}

\subsection{Alternative Candidates Capture Attention from the Fringe}

\begin{figure}[!htb] 
	\centering
		\includegraphics[width=\columnwidth]{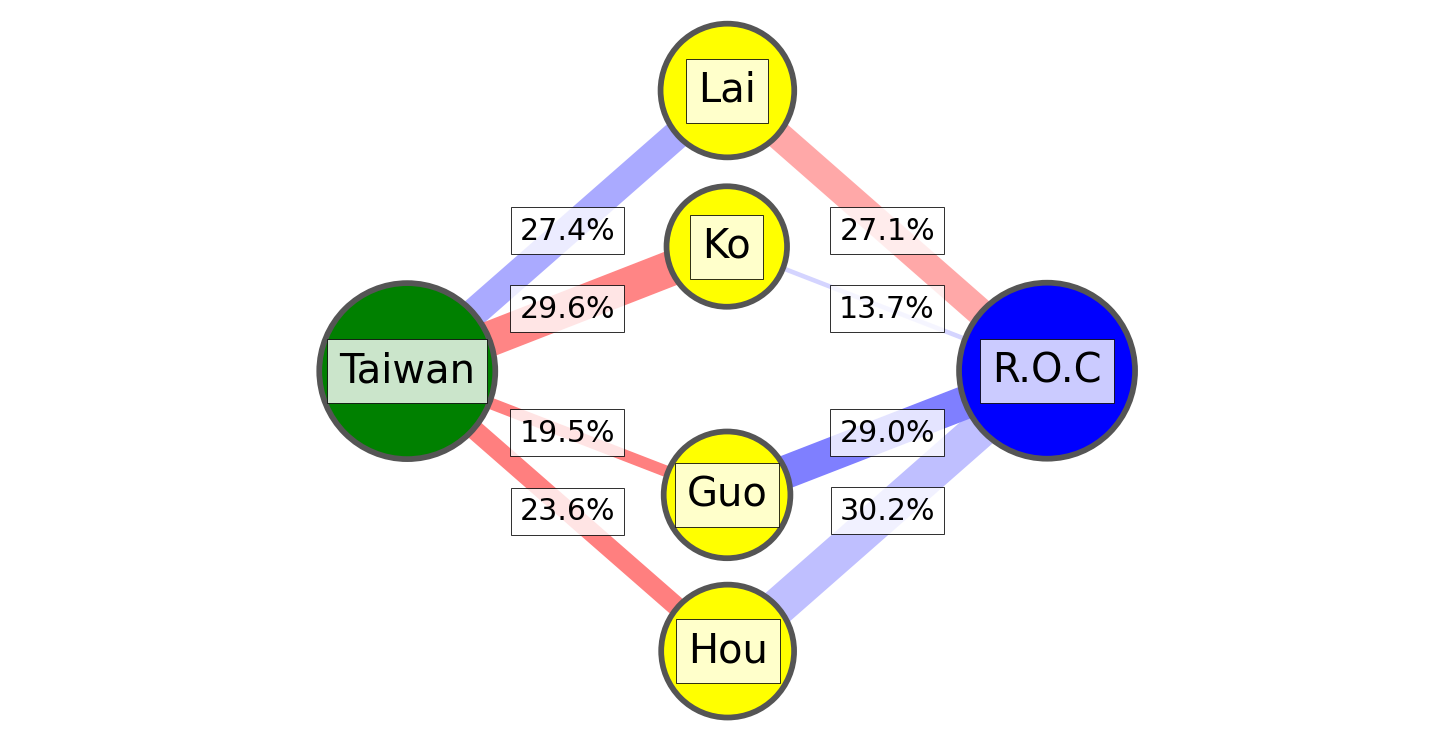}
	\caption{Groups include ones that reference ``Taiwan'' (green) and ``Republic of China'' (blue) in their name connected to presidential candidates (yellow). Edge width is the number of posts about a candidate normalized over all posts in each group set.}\label{fig:simple-cand-network}
\end{figure}

First we consider how users in Taiwan and ROC general user groups discuss the four candidates. Figure~\ref{fig:simple-cand-network} shows the network of Taiwan-affiliated public groups (green), ROC-affiliated public groups (blue), and the four candidates (yellow). The edge weights are the proportion of mentions of each candidate, by the public groups (vertically they sum to 1). The colors are the favorability toward that candidate based on Eq.~\ref{eq:love-hate-ratio}:

\begin{equation} \label{eq:love-hate-ratio}
   Fav(Cand, Groups) = \frac{| \{ \text{Love + Care Reactions}\} |}{| \{ \text{Love + Care Reactions}\} | + | \{ \text{Angry Reactions}\} |} 
\end{equation}

Here, we normalize the number of love and care reactions over the sum of the two and the number of angry reactions, which are shown from low (red) to high (blue) in the network edges of Figure~\ref{fig:simple-cand-network}. The ratio values are shown in Table~\ref{tab:edge-affect}. A full visualization containing all groups is given in the Appendix.

There are a few immediate observations. First, traditional candidates tend to be balanced in mentions in these groups. For instance, Lai is mentioned 27.1\% of the time in R.O.C groups and 27.4\% in Taiwan groups. On the other hand, Ko and Hou have a bias in mentions in each group--- Ko is mentioned 29.6\% in Taiwan groups versus 13.7\% in R.O.C groups; Hou is mentioned 29.0\% in R.O.C. groups but only 19.5\% in Taiwan groups. The conception of national identity directly plays a role in which candidates users of each group engage with.
\begin{table}[!htb]
\begin{tabular}{|l|l|l|}
\hline
\textbf{Public Group Identity} & \textbf{Candidate} & \textbf{Love-Anger Mean} \\ \hline
Taiwan                         & Lai                & 0.791                    \\ \hline
Taiwan                         & Ko                 & 0.353                    \\ \hline
Taiwan                         & Gou                & 0.342                    \\ \hline
Taiwan                         & Hou                & 0.339                    \\ \hline
R.O.C                          & Lai                & 0.428                    \\ \hline
R.O.C                          & Ko                 & 0.702                    \\ \hline
R.O.C                          & Gou                & 0.885                    \\ \hline
R.O.C                          & Hou                & 0.746                    \\ \hline
\end{tabular} \caption{Mean effect of edges between public Groups (by national identity affiliation) and presidential candidate. } \label{tab:edge-affect}
\end{table}

Additionally, the sentiment analysis shows Lai is positively-received in the Taiwan groups, whereas negatively-received in R.O.C. groups. The dynamic is flipped for Ko, Gou, and Hou, where they are positively-received in R.O.C. groups and negatively-received in Taiwan groups.  Ko is curious in that he captured the most attention amongst Taiwan groups but is negatively valenced, while is positively received in R.O.C. groups but rarely mentioned. 
The parity for these results is broken by Ko's lower ratio in Taiwan-based groups. However, this could also be due to high divisiveness across users, and a lack of younger users who prefer other platforms (i.e. Instagram).

In general, in terms of raw attention, traditional candidates receive balanced attention whereas the new candidates--- Ko and Gou--- are biased toward pan-Green and pan-Blue users respectively. This may suggest these candidates gain viability not from an undecided middle, but from the fringes on the axes of national identity. When looking at volume alone, we may posit that Ko is perceived as the largest threat from Taiwan groups and the most divisive. Most importantly, these results demonstrate that the cleavages among Taiwanese voters are not simple.

\subsection{Traditional Media supports Traditional Candidates}

\begin{figure}[htbp] 
	\centering
		\includegraphics[width=1.0\columnwidth]{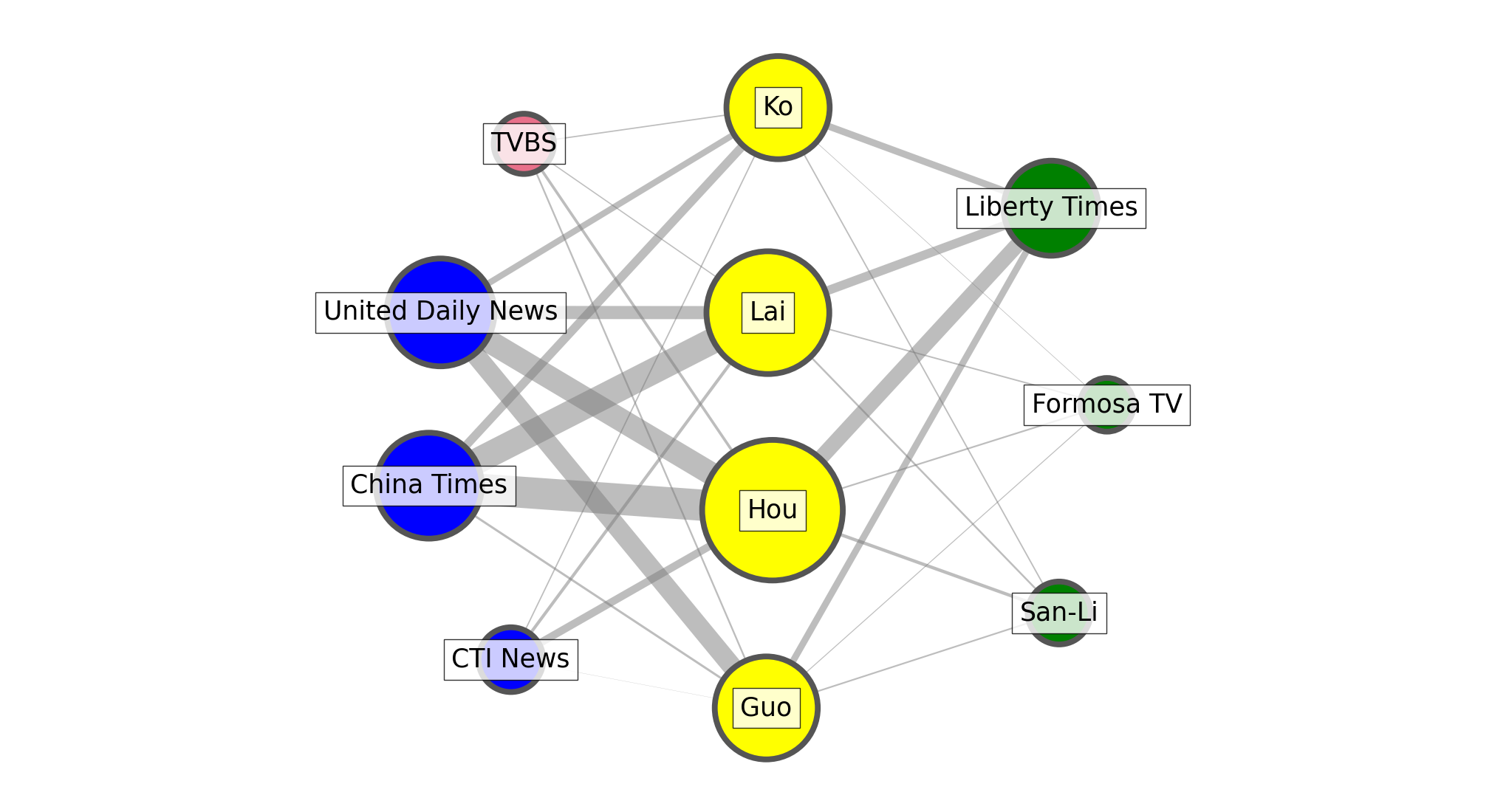}
	\caption{Network of candidate mentioning/resharing and traditional news media outlets.}\label{fig:media-network}
\end{figure}

Next, we turn our attention to the traditional media. Figure~\ref{fig:media-network} shows the number of reference made by the top seven media outlets in the dataset---Liberty Times (\zh{自由時報}), United Daily News(\zh{聯合報}), China Times(\zh{中時新聞}), CTI News(\zh{中天新聞}), Formosa TV(\zh{民視}), San-Li Media(\zh{三立電視}), and TVBS. These outlets also provide a balanced comparison: United Daily News, China Times, and CTI News are pan-Blue biased; Liberty Times, Formosa TV, and San-Li media are pan-Green biased. TVBS is considered centered~\cite{Chang_2020}.

The rise of Terry Gou and criticism toward Hou may suggest that Hou would receive the least amount of attention. In contrast, we find Hou received the most engagement, followed by Lai. Table~\ref{tab:news-media} summarizes these values and shows why this might be the case. Immediately, CTI News features Hou at 45.8\% and Gou at only 8.7\%. CTI News is known for having deep ties with the  KMT party, which may explain why they favor the traditional candidate. On the other hand, similarly blue-leaning United Daily News has a balance of coverage between Hou and Gou.

The other statistic to note is that Ko is mentioned the least by the traditional media. Although this may be deliberate, a more likely reason is that when responding to ratings, the type of comparisons that generate attention diverges based on the candidate and their background. As we will see in Section~\ref{sec:virality-reg}, candidate responses to cross-party comparisons diverge greatly. The traditional media also has a steadier diet amongst the older generation; Ko would not generate as much traction.
\begin{table}[!htb]
\begin{tabular}{|l|l|l|l|l|l|l|l|l|}
\hline
\textbf{}    & \textbf{CTI News} & \textbf{China Times} & \textbf{\begin{tabular}[c]{@{}l@{}}United \\ Daily News\end{tabular}} & \textbf{\begin{tabular}[c]{@{}l@{}}Liberty \\ Times\end{tabular}} & \textbf{\begin{tabular}[c]{@{}l@{}}San-Li\\ TV\end{tabular}} & \textbf{\begin{tabular}[c]{@{}l@{}}Formosa \\ TV\end{tabular}} & \textbf{TVBS} & \textbf{Average} \\ \hline
\textbf{Gou} & 8.7\%             & 11.0\%               & 28.8\%                                                                & 21.8\%                                                            & 24.0\%                                                       & 22.9\%                                                         & 27.5\%        & 20.7\%           \\ \hline
\textbf{Ko}  & 15.4\%            & 17.1\%               & 15.6\%                                                                & 22.3\%                                                            & 20.4\%                                                       & 18.2\%                                                         & 19.9\%        & 18.4\%           \\ \hline
\textbf{Lai} & 30.1\%            & 35.7\%               & 26.1\%                                                                & 26.4\%                                                            & 24.6\%                                                       & 31.8\%                                                         & 22.5\%        & 28.2\%           \\ \hline
\textbf{Hou} & 45.8\%            & 36.2\%               & 29.5\%                                                                & 29.5\%                                                            & 31.0\%                                                       & 27.0\%                                                         & 30.1\%        & 32.7\%           \\ \hline
\end{tabular} \caption{Proportion of attention per candidate, normalized by news media.} \label{tab:news-media}
\end{table}

Overall, traditional candidates are discussed more by the traditional media.  Certainly, this may be a factor of Lai and Hou being in the race longer. However, these results were extracted using the same direct query based on candidate name and time frame, which indicates their overall greater presence in the traditional media. Inadvertently, the institutional media may exacerbate these comparisons to their core audiences.

\subsection{Issue Salience}

\begin{figure}[htbp] 
	\centering
		\includegraphics[width=1.0\columnwidth]{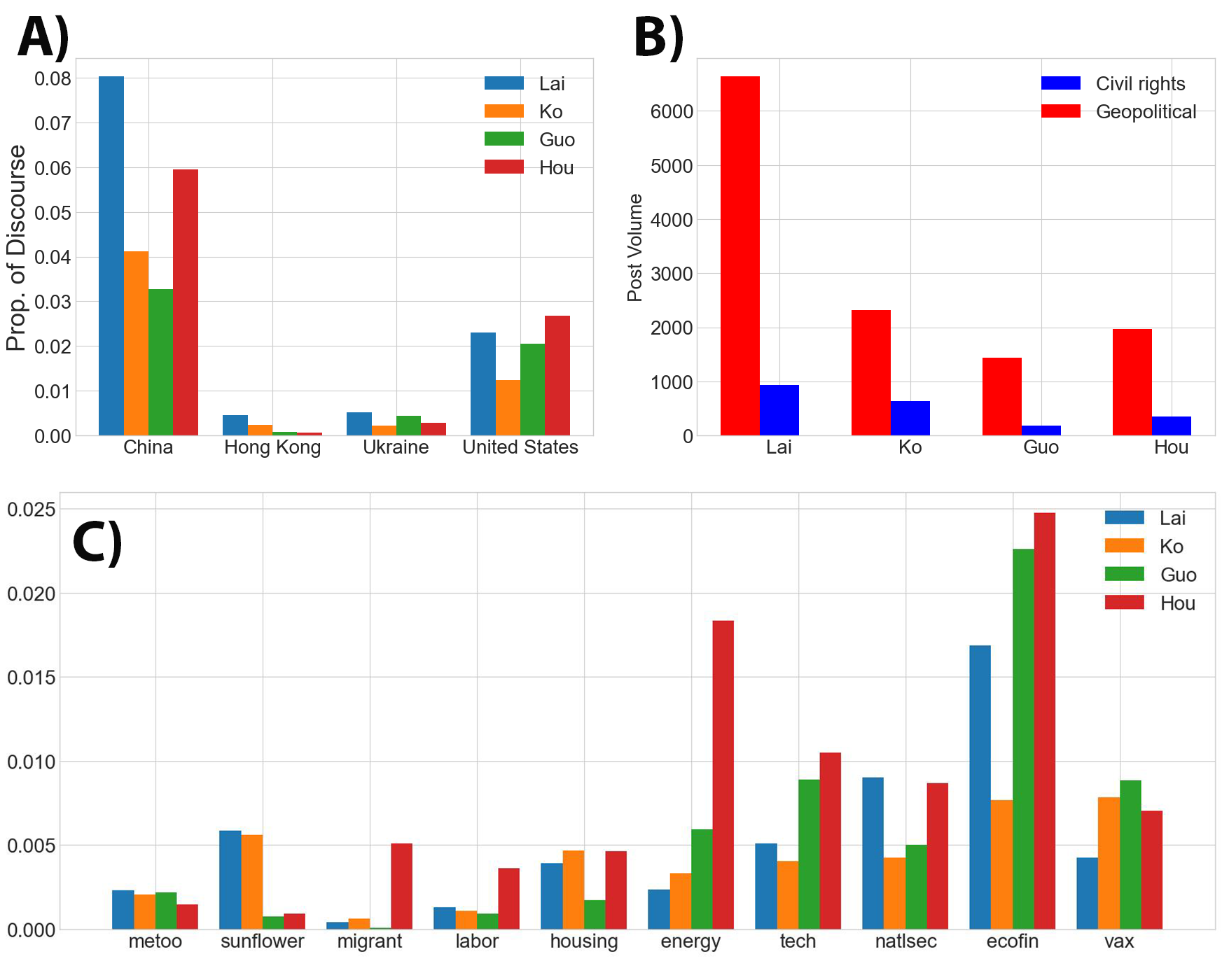}
	\caption{Volume of key topics discussed. a) Geopolitical issue proportion cross-sectioned with candidates. b) Civil rights versus geopolitical topic volume by candidate. c) Domestic policy issue discourse volume. }\label{fig:issues-bar}
\end{figure}

For this analysis, we use the posts from candidate-specific communities, to better understand which topics the supporters for each candidate discuss.   Figure~\ref{fig:issues-bar}a) shows the frequency of geopolitical issues in these support groups, and we find Lai and Hou groups having a higher proportion of discourse regarding China and the United States. On the other hand, Lai and Ko groups reference Hong Kong the most which aligns with their Pan-green identity.

Figure~\ref{fig:issues-bar}b) shows the frequency of geopolitical issues versus civil rights--- \textit{\#MeToo, the Sunflower Movement, migrant workers, labor, }and\textit{ housing.} For every candidate, the number of posts concerning geopolitics greatly outnumber civil rights. Figure~\ref{fig:issues-bar}c) shows the posts per domestic issue normalized over the total volume per candidate group. Like the trend of Hong Kong, the Sunflower Movement is evoked in Lai and Ko support groups but rarely with Gou and Hou. Hou's supporters asymmetrically discuss migrant workers and energy. 

In these supporter groups, we find traditional candidates Lai and Hou groups having a greater level of discourse regarding China and the United States, pan-Green candidates Lai and Ko groups referencing Hong Kong, and, on the flip side, pan-Blue supporters fixate more on \textit{energy}, \textit{tech}, and the\textit{ general economy}. When discussing national security, the support groups for traditional candidates (Lai and Hou) are more vocal. Lai's supporters also shy away from discussing vaccines.
\begin{figure}[htbp] 
	\centering
		\includegraphics[width=0.7\columnwidth]{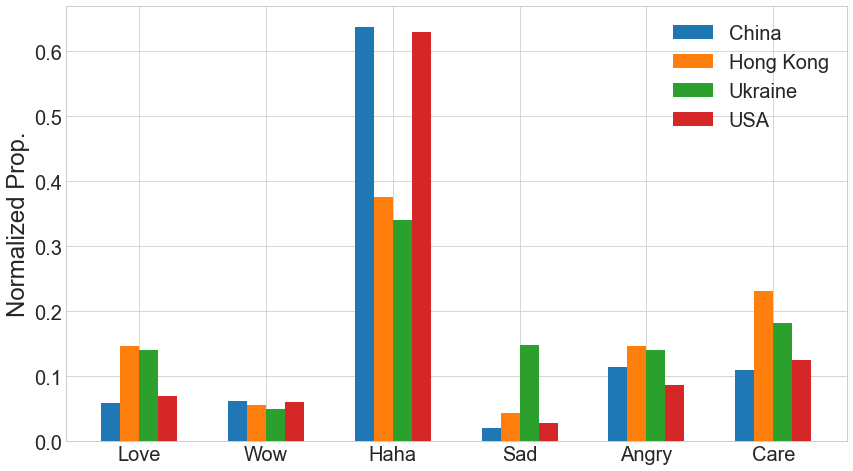}
	\caption{Geopolitical issues by affective reaction. }\label{fig:issues-affect}
\end{figure}

Given the salience of geopolitical issues, figure~\ref{fig:issues-affect} further cross sections the four geopolitical issues. Note, we forgo ''Likes'' as they account for around 90\% of all interactions. China and the USA elicit high levels of "Haha'' reactions, whereas Hong Kong and Ukraine elicit "Love'', "Angry'', and "Care'' reactions---a combination of solidarity and outrage. The invasion of Ukraine in particular elicits a significant number of ''Sad'' reactions. 

Overall, we find certain motifs in the demand of topics and candidates. Geopolitical issues concerning national security are emphasized by supporters of traditional candidates; Taiwan-based identity issues like the Sunflower Movement are emphasized by pan-Green supporters.  Macroeconomic policy like energy, tech, and the economy are emphasized by pan-Blue supporters.
While these results are descriptive, they provide an important segue to the principle results in Section~\ref{sec:virality-reg}, for predicting virality.

\subsection{Virality Regressions} \label{sec:virality-reg}

To investigate what drives virality, we regress total interactions, love reactions, and angry reactions on the salient topics and group status. Table~\ref{tab:overall-reg} shows the overall regression results for topic-based indicator variables and in-/out-group labeling.
\begin{table}[!htb]
\begin{tabular}{|l|l|l|l|}
\hline
\textbf{Topic}& \multicolumn{1}{c|}{\textbf{\begin{tabular}[c]{@{}c@{}}Total Interactions\\ (R\textasciicircum{}2 = 0.506)\end{tabular}}} & \multicolumn{1}{c|}{\textbf{\begin{tabular}[c]{@{}c@{}}Love\\ (R\textasciicircum{}2 = 0.320)\end{tabular}}} & \multicolumn{1}{c|}{\textbf{\begin{tabular}[c]{@{}c@{}}Angry\\ (R\textasciicircum{}2 = 0.361)\end{tabular}}} \\ \hline
usa      & \textbf{1.3497***}                                                                                                        & \textbf{0.7774***}                                                                                          & \textbf{0.4901***}                                                                                           \\ \hline
metoo    & 1.1299**                                                                                                                  & 0.4692*                                                                                                     & 0.5235***                                                                                                    \\ \hline
migrant  & 1.1306***                                                                                                                 & 0.5648*                                                                                                     &                                                                                                              \\ \hline
housing  & 0.811***                                                                                                                  & 0.4323***                                                                                                   & 0.2225**                                                                                                     \\ \hline
energy   & 0.6651***                                                                                                                 & 0.2854**                                                                                                    & 0.2929***                                                                                                    \\ \hline
tech     & 1.5083***                                                                                                                 & 0.7362***                                                                                                   & 0.5294***                                                                                                    \\ \hline
natlsec  & 0.7769***                                                                                                                 & 0.4104***                                                                                                   & 0.3409***                                                                                                    \\ \hline
ecofin   & 1.1731***                                                                                                                 & 0.6185***                                                                                                   & 0.4502***                                                                                                    \\ \hline
vax      & 0.876***                                                                                                                  & 0.3961**                                                                                                    & 0.4239***                                                                                                    \\ \hline
in-group& \textbf{2.8111***}                                                                                                        & \textbf{0.9878***}                                                                                          & 0.6707***                                                                                                    \\ \hline
out-group& \textbf{2.4962***}                                                                                                        & 0.5095***                                                                                                   & \textbf{0.8587***}                                                                                           \\ \hline
in\_and\_out & \textbf{-2.8169***}                                                                                                       & -1.1018***                                                                                                  & -0.8749***                                                                                                   \\ \hline
\end{tabular}\caption{Regression results on three primary engagement metrics. ***$p< 0.001$; **$p < 0.01$; *$p < 0.05$.} \label{tab:overall-reg}
\end{table}
For all three dependent variables, policy reference to the USA, MeToo, migrant workers, housing, energy, tech, national security, the economy, and vaccines all produce positive effects (except for migrant workers on Angry reacts). What is curious is the absence of China, which may suggest its ubiquity has made for no statistical increase in reactions. In contrast, foreign policy regarding the United States has a much larger impact, shadowed only by reference to Tech. These two topics often go hand-in-hand, due to the core theme of Silicon Diplomacy and debates on TSMC in Arizona.

Affect (group status) matters much more than issue topic. However, in-group favoritism seems to still overcome out-group animosity (based on the reactions).

More importantly, group labels produce the greatest effect, where in-group mentions generate a 2.81 coefficient (645  interaction increase) and out-group labels produce a 2.49 coefficient (309  interaction increase). These results stand in contrast to the findings of Rathje et al. 2021, where in-group mentions appear to be a larger driver of overall engagement (though they consider a much larger set of Congressional Members from the United States). However, we find alignment for love and angry reactions, where in-group mentions drive the love reactions, then out-group mentions drive angry reactions respectively. Additionally, when both in-and-out groups are mentioned, diffusion actually decreases, which shows direct comparisons in the same post are detrimental to diffusion.

\begin{table}[!htb]
\begin{tabular}{|lllll|}
\hline
\multicolumn{5}{|c|}{Total Interaction}                                                                                                                                                                                                                                                                                                                                                                                                       \\ \hline
\multicolumn{1}{|l|}{}             & \multicolumn{1}{c|}{\begin{tabular}[c]{@{}c@{}}Ko\\ (R\textasciicircum{}2 = 0.387)\end{tabular}} & \multicolumn{1}{c|}{\begin{tabular}[c]{@{}c@{}}Hou \\ (R\textasciicircum{}2 = 0.597)\end{tabular}} & \multicolumn{1}{c|}{\begin{tabular}[c]{@{}c@{}}Lai \\ (R\textasciicircum{}2 = 0.521)\end{tabular}} & \multicolumn{1}{c|}{\begin{tabular}[c]{@{}c@{}}Gou \\ (R\textasciicircum{}2 = 0.455)\end{tabular}} \\ \hline
\multicolumn{1}{|l|}{china}    & \multicolumn{1}{l|}{}                                                                            & \multicolumn{1}{l|}{1.2155*}                                                                       & \multicolumn{1}{l|}{}                                                                              &                                                                                                    \\ \hline
\multicolumn{1}{|l|}{usa}      & \multicolumn{1}{l|}{1.8061***}                                                                   & \multicolumn{1}{l|}{1.7916**}                                                                      & \multicolumn{1}{l|}{}                                                                              & 1.839**                                                                                            \\ \hline
\multicolumn{1}{|l|}{metoo}    & \multicolumn{1}{l|}{}                                                                            & \multicolumn{1}{l|}{2.59*}                                                                         & \multicolumn{1}{l|}{}                                                                              &                                                                                                    \\ \hline
\multicolumn{1}{|l|}{housing}  & \multicolumn{1}{l|}{1.8713**}                                                                    & \multicolumn{1}{l|}{}                                                                              & \multicolumn{1}{l|}{}                                                                              &                                                                                                    \\ \hline
\multicolumn{1}{|l|}{energy}   & \multicolumn{1}{l|}{1.8664**}                                                                    & \multicolumn{1}{l|}{}                                                                              & \multicolumn{1}{l|}{}                                                                              &                                                                                                    \\ \hline
\multicolumn{1}{|l|}{tech}     & \multicolumn{1}{l|}{1.8272**}                                                                    & \multicolumn{1}{l|}{1.6557*}                                                                       & \multicolumn{1}{l|}{1.3971***}                                                                     & 1.5704*                                                                                            \\ \hline
\multicolumn{1}{|l|}{ecofin}   & \multicolumn{1}{l|}{1.2597*}                                                                     & \multicolumn{1}{l|}{}                                                                              & \multicolumn{1}{l|}{1.4511***}                                                                     & 1.1469*                                                                                            \\ \hline
\multicolumn{1}{|l|}{in\_group}       & \multicolumn{1}{l|}{3.0881***}                                                                   & \multicolumn{1}{l|}{3.6666***}                                                                     & \multicolumn{1}{l|}{2.9961***}                                                                     & 3.0399*                                                                                            \\ \hline
\multicolumn{1}{|l|}{out\_group}      & \multicolumn{1}{l|}{2.6804***}                                                                   & \multicolumn{1}{l|}{3.4528***}                                                                     & \multicolumn{1}{l|}{}                                                                              & 2.5003***                                                                                          \\ \hline
\multicolumn{1}{|l|}{in\_and\_out} & \multicolumn{1}{l|}{-3.6866**}                                                                   & \multicolumn{1}{l|}{-3.7982***}                                                                    & \multicolumn{1}{l|}{}                                                                              & -4.1783*                                                                                           \\ \hline
\end{tabular} \caption{Candidate-level regressions on total interactions.***$p< 0.001$; **$p < 0.01$; *$p < 0.05$.} \label{tab:total-interaction}
\end{table}

Next, we conduct the same regression for each of the four candidates individually, shown in Tab~\ref{tab:total-interaction}.  Once more, we find in-group mentions are crucial to driving diffusion. There are a few trends to note. First, Hou's community is the only one who elicits a reaction in mentioning China. Although this may seem strange, one explanation is that China is such a ubiquitous topic of comparison, that it does not generate substantially more diffusion among the communities of other candidates. Other issues such as the USA, defense, and technology may all reference China---and hence act as an invisible confounder. On the contrary, the keywords for Hou's China include significant reference to better economic ties. As expected, Hou is also missing in significance for the general economy, because his economic discourse is tied heavily in cross-strait relations.

Curiously, only Lai does not elicit more responses when mentioning the United States. Moreover, his recipe for total engagement is simple--- just three indicator columns generate statistically significant results. The absence of out-group mentions and in-/out-group comparisons suggests his supporter base may be agnostic to out-group animosity.

The regression results for Total Interaction align with the love reaction regression (shown in the Appendix). However, we see a different story when regression on the Angry reaction, shown in Table~\ref{tab:angry-reg}. While the regression variables remain roughly the same for Ko and Hou, there is a dramatic increase in significant variables for Lai and a decrease for Gou. Lai adds on the USA, MeToo, labor and wages, housing, tech, and the economy; Gou loses Tech and the Economy.

\begin{table}[!htb]
\begin{tabular}{|lllll|}
\hline
\multicolumn{5}{|c|}{Angry Reaction}                                                                                                                                                                                                                                                                                                                                                                                                                  \\ \hline
\multicolumn{1}{|c|}{}             & \multicolumn{1}{c|}{\begin{tabular}[c]{@{}c@{}}Ko \\ (R\textasciicircum{}2 = 0.348)\end{tabular}} & \multicolumn{1}{c|}{\begin{tabular}[c]{@{}c@{}}Hou \\ (R\textasciicircum{}2 = 0.570)\end{tabular}} & \multicolumn{1}{c|}{\begin{tabular}[c]{@{}c@{}}Lai \\ (R\textasciicircum{}2 = 0.503)\end{tabular}} & \multicolumn{1}{c|}{\begin{tabular}[c]{@{}c@{}}Gou \\ (R\textasciicircum{}2 = 0.421)\end{tabular}} \\ \hline
\multicolumn{1}{|l|}{china}    & \multicolumn{1}{l|}{}                                                                             & \multicolumn{1}{l|}{0.7396**}                                                                      & \multicolumn{1}{l|}{}                                                                              &                                                                                                    \\ \hline
\multicolumn{1}{|l|}{usa}      & \multicolumn{1}{l|}{0.4344*}                                                                      & \multicolumn{1}{l|}{0.6445*}                                                                       & \multicolumn{1}{l|}{0.2518*}                                                                       & 0.5145**                                                                                           \\ \hline
\multicolumn{1}{|l|}{metoo}    & \multicolumn{1}{l|}{}                                                                             & \multicolumn{1}{l|}{1.2127**}                                                                      & \multicolumn{1}{l|}{0.8513*}                                                                       &                                                                                                    \\ \hline
\multicolumn{1}{|l|}{labor}    & \multicolumn{1}{l|}{}                                                                             & \multicolumn{1}{l|}{}                                                                              & \multicolumn{1}{l|}{0.7978**}                                                                      &                                                                                                    \\ \hline
\multicolumn{1}{|l|}{housing}  & \multicolumn{1}{l|}{0.707**}                                                                      & \multicolumn{1}{l|}{}                                                                              & \multicolumn{1}{l|}{0.4716**}                                                                      &                                                                                                    \\ \hline
\multicolumn{1}{|l|}{energy}   & \multicolumn{1}{l|}{0.6204**}                                                                     & \multicolumn{1}{l|}{}                                                                              & \multicolumn{1}{l|}{}                                                                              &                                                                                                    \\ \hline
\multicolumn{1}{|l|}{tech}     & \multicolumn{1}{l|}{0.6508**}                                                                     & \multicolumn{1}{l|}{0.622*}                                                                        & \multicolumn{1}{l|}{0.5151***}                                                                     &                                                                                                    \\ \hline
\multicolumn{1}{|l|}{ecofin}   & \multicolumn{1}{l|}{0.4095*}                                                                      & \multicolumn{1}{l|}{}                                                                              & \multicolumn{1}{l|}{0.3548***}                                                                     &                                                                                                    \\ \hline
\multicolumn{1}{|l|}{in\_group}       & \multicolumn{1}{l|}{1.0836***}                                                                    & \multicolumn{1}{l|}{1.5249***}                                                                     & \multicolumn{1}{l|}{0.8815***}                                                                     & 0.9167*                                                                                            \\ \hline
\multicolumn{1}{|l|}{out\_group}      & \multicolumn{1}{l|}{1.2054***}                                                                    & \multicolumn{1}{l|}{1.4727***}                                                                     & \multicolumn{1}{l|}{}                                                                              & 0.9444***                                                                                          \\ \hline
\multicolumn{1}{|l|}{in\_and\_out} & \multicolumn{1}{l|}{-1.2811***}                                                                   & \multicolumn{1}{l|}{-1.2272**}                                                                     & \multicolumn{1}{l|}{}                                                                              & -1.2203*                                                                                           \\ \hline
\end{tabular} \caption{Candidate-level regression on Angry Reactions.} \label{tab:angry-reg}
\end{table}

 While it is uncertain whether this is out-group reactions to his post, another plausible explanation is that Lai elicits targeted outrage for key issues---the USA, MeToo, Labor, Housing, Tech, and the Economy. The topics that did not generate any statistical significance for general interaction, or suddenly extremely salient, when producing the angry reaction. Moreover, references to the out-group and in-/out-group comparisons remain non-influential to diffusion.

\begin{figure}[htbp] 
	\centering
		\includegraphics[width=0.9\columnwidth]{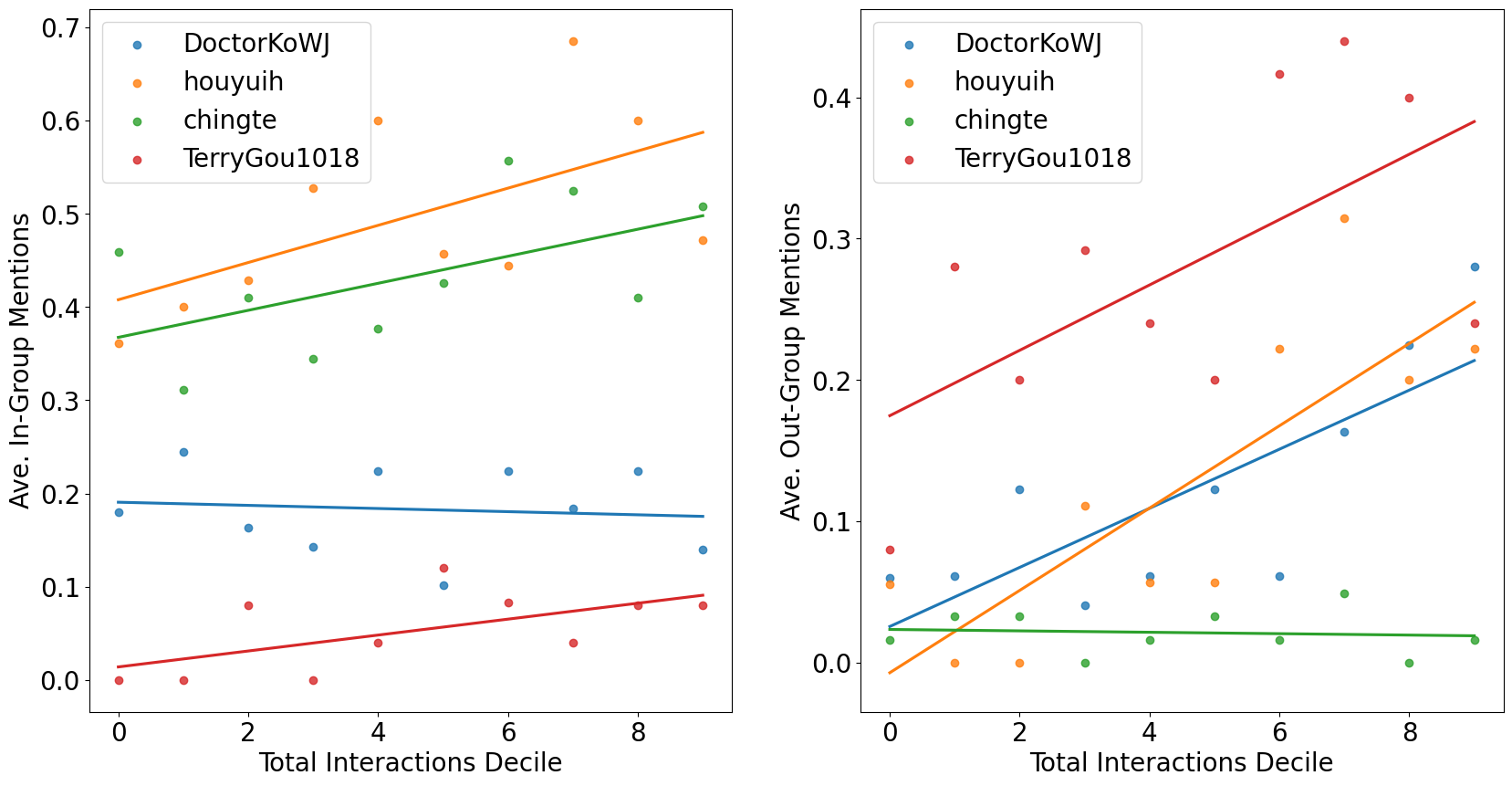}
	\caption{Regressions by candidate, by interaction decile against a) in-group mentions and b) out-group mentions.}\label{fig:reg-in-out}
\end{figure}

We explore the effects of group dynamics further in Figure~\ref{fig:reg-in-out}. Figure~\ref{fig:reg-in-out}a) shows the effects of mentioning the in-group on virality. Both traditional candidates Lai and Hou generate increase in interactions as in-group mentions increase. This is likely due to playing to their existing voter base and party legacy. Gou has overall less in-group mentions but also generates an increase of interaction.  However for Ko, when in-group mentions increase there is a slight decrease for total interactions. This suggests Ko generates virality by making out-group comparisons.

Figure~\ref{fig:reg-in-out}b) shows the effects of mentioning the out-group on virality.  Here, we find Gou having the most out-group mentions, whereas Ko and Hou are roughly the same. Virality increases for Hou, Ko, and Gou as out-group mentions increase. However, Lai's virality remains agnostic to out-group mentions.

\begin{figure}[htbp] 
	\centering
		\includegraphics[width=1.0\columnwidth]{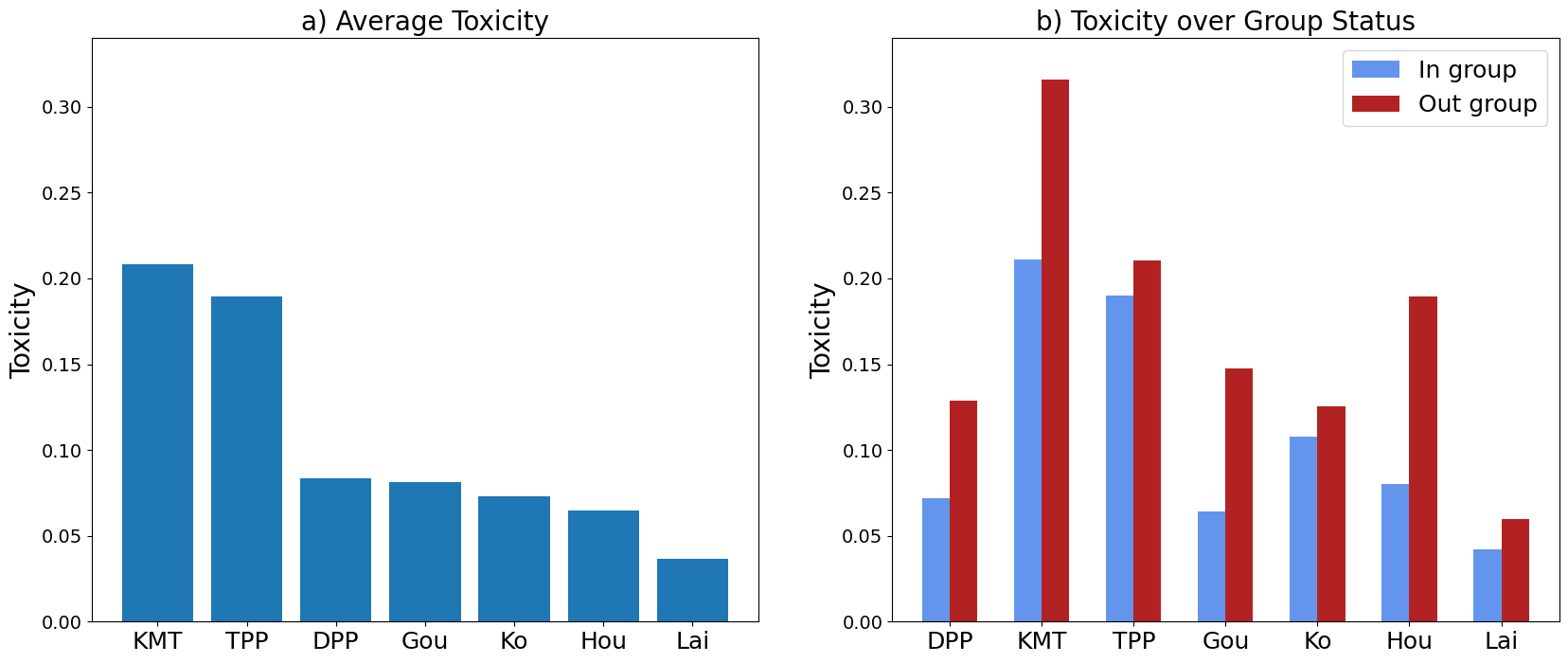}
	\caption{Average toxicity by candidate for a) overall posts and b) in-group and out-group references.}\label{fig:toxicity}
\end{figure}

Figure~\ref{fig:toxicity}a) shows the toxicity scores for all candidates, where we find the KMT and TPP with the highest scores, then Gou, Ko, Hou, and Lai in order. In general, party toxicity scores are much higher than their respective candidates. Figure~\ref{fig:toxicity}b) shows in-/out-group status cross-sectioned with toxicity scores. In alignment with the literature, out-group mentions are higher than in-group mentions, and this is particularly exacerbated for the KMT. Hou emerges as having the greatest out-group toxicity, likely due to all the comparisons he has to make--- not just with Lai and Ko, but also a contender within his own party (as Gou is coded as an out-group for the case of Hou). The low toxicity scores of the DPP and Lai reflect the non-significance of out-group references for Lai.

\section{Discussion}

In this paper, we investigate the supply-demand dynamics of the four presidential candidates of Taiwan. In comparing traditional and alternative candidates, we find traditional candidates generate more engagement from the traditional media, converge on geopolitical issues more, and are more central in terms of national identity. This last point is perhaps surprising; according to the missing middle hypothesis, we would assume the alternative candidates would grasp the vote of moderate independents. However, our results suggest that independent moderates who engage with more diverse issues simultaneously contribute to the decoupling of national identity with partisan identity. For instance, Ko generates the greatest response amongst young people who also have the greatest sense of Taiwanese identity. Similarly, Gou speaks to older, pan-Blue voters often from military families. In other words, alternative candidates gain viability from the fringes rather than the center, relative to national identity. We discuss this in detail later.

We also find the traditional media supports traditional candidates, generating 10-12\% more attention to Lai and Hou via resharing within candidate support groups. In particular, there is evidence that outlets such as CTI news seems to suppress news about Gou, relative to Hou. Discourse on geopolitics converges more heavily on traditional candidates and vastly outnumbers civil rights discourse, which may tilt the balance toward Lai and Hou.

Regarding virality, we find in-group mentions drive greater engagement rather than out-group comparisons. This diverges from Rathje et al. (2021) that finds out-group animosity generating the most virality. Although one explanation may be due to a difference in our unit of analysis---Members of Congress versus Presidential Candidates--- we could expect the level of animosity during presidential elections to be higher. This is substantiated by the alignment in the finer details: in-group references better predict positive affect whereas out-group references produce negative affect. Otherwise, this makes Taiwan an interesting case-study to understand conditions for which in-group favoritism dominates out-group animosity.

Comparing Lai and Ko, Lai's engagement is characterized by in-group references and targeted issue-based outrage, whereas Ko's engagement is characterized by out-group references. While Ko is known for advocating for domestic issues (as evidenced by the suite of statistically significant issues), he also must draw comparisons to other candidates and does not seem to generate traction by appealing to party identity. His interactions do not increase with in-group mentions. On the other hand, the two traditional candidates can generate virality by appealing to their party identity. Although Lai's lack of interaction relative to out-group mentions might be because the DPP is the incumbent party--- hence the TPP and KMT attack it--- the low toxicity levels for Lai indicate the lack of comparisons may be one of his strategies, and thus least polarizing.

Third, and perhaps most importantly, China does not predict strong virality. Although it is tempting to say this diverges from prior elections, what we observe is the issue salience is replaced instead by references to the United States and technology. Given the importance of  "Chip Wars" and silicon diplomacy to Taiwan, this indicates a shift in how the Taiwanese perceive themselves within geopolitical debates.
Although candidate viability is a core issue, it falls out of the scope of our study, though viability comparisons based on the incumbent's gender is a rich area of research~\cite{chang2023will}. Foreign interference and misinformation is yet another topic worthy of further investigation.

Our results align with the largest survey (n=15,000) conducted by Commonwealth Magazine, with additional dynamics~\cite{Cheng_2023}. First, the survey reports decreasing interest in the 92 Consensus and cross-strait relationships, and an increase toward foreign relations. As found from our regression results, Taiwan's relationship with the United States stands as one of the most crucial issues and China  does not influence engagement. In other words, although the focus has shifted to international relations and security commitments, this does not make cross-strait relations any less important. Indeed, when observing the overall volume of attention, geopolitical issues greatly outnumber domestic issues. While there are certainly concerns of the representational bias on social media toward actual voters, Taiwan, relative to other countries, has a relatively high and concentrated level of Facebook use. One small point of contention is that although there is a shift toward domestic issues, geopolitical issues still overwhelm domestic ones. In general, both the survey and our study find cross-strait geopolitical tensions are recast as an international one. 

Additionally, only Hou's virality is positively correlated with China and missing from the general economic discourse. In tandem with the survey this makes sense--- KMT supporters are the only ones who care more about cross-strait economic development than the DPP, TPP, and nonpartisan voters. In other words, Hou's discourse is inadvertently wrapped up in economic relations with China. 

\begin{figure}[htbp] 
	\centering
		\includegraphics[width=0.5\columnwidth]{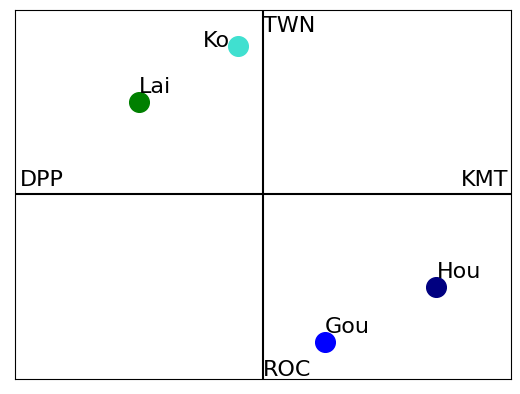}
	\caption{Conceptual model of the misalignment of partisan and national identity.}\label{fig:misalignment}
\end{figure}

Perhaps the largest divergence lies in "the missing middle" argument, which states moderate independents tend to be issue driven and seek alternative candidates. What we find is the opposite--- it is in fact those with a more radical notion of national identity that is seeking alternative candidates. How can this make sense? Ko's largest number of supporters arise from young voters, who are also the biggest proponents of Taiwanese national identity. Similarly, Gou captures older, R.O.C. voters. We illustrate this in Figure~\ref{fig:misalignment}. As such, it will be interesting to see if age can serve as a good proxy for national identity. Although moderation should be relative to national identity, the decoupling of national and partisan identity and issue ownership mean moderate independents should be measured against partisan identity, at least in this election. This band of instability, though narrow, is where independents can gain strength.

The definition of moderate has shifted. This election is significant not in the novel emergence of four candidates, but in the conditions that produced their viability. As China grew more authoritarian and in the aftermath of the Sunflower Movement and anti-extradition and National Security law in Hong Kong, there has been a gradual collapse of R.O.C. identity as defined with the China issue. This misaligned the Taiwan / R.O.C dichotomy and the DPP / KMT split, which allowed new candidates to emerge and gain ownership over new issues. 

It brings up interesting possibilities in comparative contexts. What if there is a shock such that ideological identity misaligns with partisan identity? Perhaps then, we can observe an unfreezing across deep, partisan lines.

\section*{Acknowledgments}
H.C. would like to thank Tom Hollihan, Adrian Rauschfleish, and James Druckman for their comments.





\printbibliography

@misc{Economist_2021, title={The most dangerous place on Earth}, url={https://www.economist.com/leaders/2021/05/01/the-most-dangerous-place-on-earth}, journal={The Economist}, publisher={The Economist Newspaper}, author={The Economist}, year={2021}, month={May}}

@misc{Smith_2023, title={Taiwan People's Party takes stance on "one China" and "1992 consensus"}, url={https://www.taiwannews.com.tw/en/news/4870837}, journal={Taiwan News}, publisher={Taiwan News}, author={Smith, Courtney Donovan}, year={2023}, month={Apr}}

@misc{Wei_2023, title={How a netflix show sparked a \#MeToo wave in Taiwan}, url={https://foreignpolicy.com/2023/08/12/wave-makers-netflix-taiwan-metoo-sexual-harassment-dpp-china/}, journal={Foreign Policy}, publisher={Foreign Policy}, author={Wei, Clarissa}, year={2023}, month={Aug}}

@misc{jieba,
  author = {Sun, Andy and \zhs{项超} and Schaaf,Herman and Lin, Zhe and Yan, Felix},
  title = {jieba},
  year = {2013},
  publisher = {GitHub},
  journal = {GitHub repository},
  howpublished = {\url{https://github.com/LiveMirror/jieba}},
  commit = {b050bfe}
}

@misc{chinese_converter,
  author = {Juang, Zachary and Blancada, Victor Angelo},
  title = {chinese\_converter},
  year = {2023},
  publisher = {GitHub},
  journal = {GitHub repository},
  howpublished = {\url{https://github.com/zachary822/chinese-converter}},
  commit = {ddc46d7}
}

@article{chang2022justiceforgeorgefloyd,
  title={\# JusticeforGeorgeFloyd: How Instagram facilitated the 2020 Black Lives Matter protests},
  author={Chang, Ho-Chun Herbert and Richardson, Allissa and Ferrara, Emilio},
  journal={PLoS one},
  volume={17},
  number={12},
  pages={e0277864},
  year={2022},
  publisher={Public Library of Science San Francisco, CA USA}
}

@book{miller2022chip,
  title={Chip war: the fight for the world's most critical technology},
  author={Miller, Chris},
  year={2022},
  publisher={Simon and Schuster}
}

@book{achen2017taiwan,
  title={The Taiwan Voter},
  author={Achen, Christopher and Wang, Te-Yu},
  year={2017},
  publisher={University of Michigan Press}
}

@article{chang2023liberals,
  title={Liberals Engage with More Diverse Policy Topics and Toxic Content Than Conservatives on Social Media},
  author={Chang, Ho-Chun Herbert and Druckman, James and Ferrara, Emilio and Willer, Robb},
  year={2023},
  publisher={OSF Preprints}
}

@article{chang2023will,
  title={Will She Win? Gendered Media Coverage of the 2020 Democratic Party Presidential Primaries},
  author={Chang, Ho-Chun Herbert and Brichta, Maximilian and Ahn, Soyun and de Vight, Jackson},
  journal={Journalism \& Mass Communication Quarterly},
  pages={10776990231182154},
  year={2023},
  publisher={SAGE Publications Sage CA: Los Angeles, CA}
}

@article{jacoby2010policy,
  title={Policy attitudes, ideology and voting behavior in the 2008 election},
  author={Jacoby, William G},
  journal={Electoral Studies},
  volume={29},
  number={4},
  pages={557--568},
  year={2010},
  publisher={Elsevier}
}

@article{linz1967party,
  title={Party Systems and Voter Alignments},
  author={Linz, J and Lipset, Seymour M and Rokkan, Stein},
  year={1967}
}

@article{goel2016structural,
  title={The structural virality of online diffusion},
  author={Goel, Sharad and Anderson, Ashton and Hofman, Jake and Watts, Duncan J},
  journal={Management Science},
  volume={62},
  number={1},
  pages={180--196},
  year={2016},
  publisher={INFORMS}
}

@article{wang2004contending,
  title={Contending identities in Taiwan: Implications for cross-strait relations},
  author={Wang, Te-Yu and Liu, I-Chou},
  journal={Asian Survey},
  volume={44},
  number={4},
  pages={568--590},
  year={2004},
  publisher={JSTOR}
}

@incollection{corcuff2016taiwan,
  title={Taiwan's "Mainlanders," New Taiwanese?},
  author={Corcuff, St{\'e}phane},
  booktitle={Memories of the Future},
  pages={163--195},
  year={2016},
  publisher={Routledge}
}

@article{liu1999taiwanese,
  title={The Taiwanese/Chinese identity of the Taiwan people},
  author={Liu, I-Chou and Ho, Szu-Yin},
  journal={Issues \& Studies},
  volume={35},
  number={3},
  pages={1--34},
  year={1999},
 publisher={dddd}
}

@article{triandafyllidou1998national,
  title={National identity and the'other'},
  author={Triandafyllidou, Anna},
  journal={Ethnic and racial studies},
  volume={21},
  number={4},
  pages={593--612},
  year={1998},
  publisher={Taylor \& Francis}
}

@article{yu2017parties,
  title={Parties, Partisans, and Independents in Taiwan},
  author={Yu, Ching-hsin},
  journal={The Taiwan Voter},
  pages={71},
  year={2017},
  publisher={University of Michigan Press Ann Arbor}
}

@article{gidron2022many,
  title={Many ways to be right: cross-pressured voters in Western Europe},
  author={Gidron, Noam},
  journal={British Journal of Political Science},
  volume={52},
  number={1},
  pages={146--161},
  year={2022},
  publisher={Cambridge University Press}
}

@book{stonecash2018diverging,
  title={Diverging parties: Social change, realignment, and party polarization},
  author={Stonecash, Jeff},
  year={2018},
  publisher={Routledge}
}

@article{abramowitz1998ideological,
  title={Ideological realignment in the US electorate},
  author={Abramowitz, Alan I and Saunders, Kyle L},
  journal={The Journal of Politics},
  volume={60},
  number={3},
  pages={634--652},
  year={1998},
  publisher={University of Texas Press}
}

@book{shafer1991end,
  title={The End of Realignment?: Interpreting American Electoral Eras},
  author={Shafer, Byron E},
  year={1991},
  publisher={Univ of Wisconsin Press}
}

@article{wang2017changing,
  title={Changing Boundaries},
  author={Wang, TY},
  journal={The Taiwan Voter},
  pages={45},
  year={2017},
  publisher={University of Michigan Press}
}

@article{sheng2017issues,
  title={Issues, political cleavages, and party competition in Taiwan},
  author={Sheng, Shing-yuan and Liao, Hsiao-chuan Mandy},
  journal={The Taiwan Voter},
  pages={98},
  year={2017},
  publisher={University of Michigan Press Ann Arbor}
}

@article{sheng2013issues,
  title={Issues, Party Performance and Voters' Voting Behavior},
  author={Sheng, Shing-yuan},
  journal={2012 Presidential and Legislative Elections: Changes and Continuation},
  pages={203--27},
  year={2013}
}

@article{stokes1963spatial,
  title={Spatial models of party competition},
  author={Stokes, Donald E},
  journal={American political science review},
  volume={57},
  number={2},
  pages={368--377},
  year={1963},
  publisher={Cambridge University Press}
}

@article{lin2011cross,
  title={Cross-strait Trade and Class Politics in Taiwan.},
  author={Lin, Thung-Hong and Hu, Alfred Ko-We},
  journal={Si yu Yan (Thought \& Words)},
  volume={49},
  number={3},
  year={2011}
}

@article{lee2017we,
  title={Are we rational or not? The exploration of voter choices during the 2016 presidential and legislative elections in Taiwan},
  author={Lee, I-Ching and Chen, Eva E and Yen, Nai-Shing and Tsai, Chia-Hung and Cheng, Hsu-Po},
  journal={Frontiers in Psychology},
  volume={8},
  pages={1762},
  year={2017},
  publisher={Frontiers Media SA}
}

@article{wang2019myth,
    title={The myth of polarization among Taiwanese voters: The missing middle},
    author={Wang, Austin Horng-En},
    journal={Journal of East Asian Studies},
    volume={19},
    number={3},
    pages={275--287},
    year={2019},
    publisher={Cambridge University Press}
}

@article{frimer2023incivility,
  title={Incivility is rising among American politicians on Twitter},
  author={Frimer, Jeremy A and Aujla, Harinder and Feinberg, Matthew and Skitka, Linda J and Aquino, Karl and Eichstaedt, Johannes C and Willer, Robb},
  journal={Social Psychological and Personality Science},
  volume={14},
  number={2},
  pages={259--269},
  year={2023},
  publisher={SAGE Publications Sage CA: Los Angeles, CA}
}

@article{wang2023does,
    title={Why Does Taiwan Identity Decline?},
    author={Wang, Austin Horng-En and Yeh, Yao-Yuan and Wu, Charles KS and Chen, Fang-Yu},
    journal={Journal of Asian and African Studies},
    pages={00219096231168068},
    year={2023},
    publisher={SAGE Publications Sage UK: London, England}
}

@article{rathje2021out,
  title={Out-group animosity drives engagement on social media},
  author={Rathje, Steve and Van Bavel, Jay J and Van Der Linden, Sander},
  journal={Proceedings of the National Academy of Sciences},
  volume={118},
  number={26},
  pages={e2024292118},
  year={2021},
  publisher={National Acad Sciences}
}

@misc{Cheng_2023, title={2024 Presidential Elections Decoded: Near-half of Citizens wary of war and atrophy of the 92 Consensus (\zh{2024總統大選，最大規模民調解密：近半民眾憂開戰、「九二共識」市場萎縮)}}, url={https://www.cw.com.tw/article/5127580}, journal={Commonwealth Magazine (\zh{天下雜誌)}}, publisher={Commonwealth Magazine (\zh{天下雜誌})}, author={Cheng, Min-Sheng}, year={2023}, month={Oct}}

@book{levendusky2009partisan,
  title={The partisan sort: How liberals became Democrats and conservatives became Republicans},
  author={Levendusky, Matthew},
  year={2009},
  publisher={University of Chicago Press}
}

@article{blei2003latent,
  title={Latent dirichlet allocation},
  author={Blei, David M and Ng, Andrew Y and Jordan, Michael I},
  journal={Journal of machine Learning research},
  volume={3},
  number={Jan},
  pages={993--1022},
  year={2003}
}

@article{chang2021digital,
  title={Digital civic participation and misinformation during the 2020 Taiwanese presidential election},
  author={Chang, Ho-Chun Herbert and Haider, Samar and Ferrara, Emilio},
  journal={Media and Communication},
  volume={9},
  number={1},
  pages={144--157},
  year={2021}
}

@article{huszar2022algorithmic,
  title={Algorithmic amplification of politics on Twitter},
  author={Huszar, Ferenc and Ktena, Sofia Ira and O'Brien, Conor and Belli, Luca and Schlaikjer, Andrew and Hardt, Moritz},
  journal={Proceedings of the National Academy of Sciences},
  volume={119},
  number={1},
  pages={e2025334119},
  year={2022},
  publisher={National Acad Sciences}
}

@article{kingzette2021affective,
  title={How affective polarization undermines support for democratic norms},
  author={Kingzette, Jon and Druckman, James N and Klar, Samara and Krupnikov, Yanna and Levendusky, Matthew and Ryan, John Barry},
  journal={Public Opinion Quarterly},
  volume={85},
  number={2},
  pages={663--677},
  year={2021},
  publisher={Oxford University Press}
}

@article{guess2023reshares,
  title={Reshares on social media amplify political news but do not detectably affect beliefs or opinions},
  author={Guess, Andrew M and Malhotra, Neil and Pan, Jennifer and Barber{\'a}, Pablo and Allcott, Hunt and Brown, Taylor and Crespo-Tenorio, Adriana and Dimmery, Drew and Freelon, Deen and Gentzkow, Matthew and others},
  journal={Science},
  volume={381},
  number={6656},
  pages={404--408},
  year={2023},
  publisher={American Association for the Advancement of Science}
}

@article{nyhan2023like,
  title={Like-minded sources on Facebook are prevalent but not polarizing},
  author={Nyhan, Brendan and Settle, Jaime and Thorson, Emily and Wojcieszak, Magdalena and Barber{\'a}, Pablo and Chen, Annie Y and Allcott, Hunt and Brown, Taylor and Crespo-Tenorio, Adriana and Dimmery, Drew and others},
  journal={Nature},
  pages={1--8},
  year={2023},
  publisher={Nature Publishing Group UK London}
}

@article{hollihan2023public,
  title={Public Diplomacy Arguments and Taiwan},
  author={Hollihan, Thomas A and Riley, Patricia},
  journal={Journal of Public Diplomacy Vol},
  volume={3},
  number={1},
  pages={14--36},
  year={2023}
}

@article{nachman2023routine,
  title={Routine Problems: Movement Party Institutionalization and the Case of Taiwan's New Power Party},
  author={Nachman, Lev},
  journal={Studies in Comparative International Development},
  pages={1--20},
  year={2023},
  publisher={Springer}
}

@article{rigger2022unification,
  title={Why is unification so unpopular in Taiwan? Its the PRC political system, not just culture},
  author={Rigger, Shelley and Nachman, Lev and Mok, Chit Wai John and Chan, Nathan Kar Ming},
  year={2022},
  publisher={Brookings Institution}
}

@article{rigger2021people,
  title={How are people feeling in the most dangerous place on Earth?},
  author={Rigger, Shelley and Nachman, Lev and Mok, Chit Wai John and Chan, Nathan Kar Ming},
  year={2021},
  publisher={Brookings Institution}
}

@article{druckman2019we,
  title={What do we measure when we measure affective polarization?},
  author={Druckman, James N and Levendusky, Matthew S},
  journal={Public Opinion Quarterly},
  volume={83},
  number={1},
  pages={114--122},
  year={2019},
  publisher={Oxford University Press UK}
}

@article{ortiz2023rise,
  title={The rise of social media},
  author={Ortiz-Ospina, Esteban and Roser, Max},
  journal={Our world in data},
  year={2023}
}

@misc{Chang_2020, title={Covid-19 media reporting in Taiwan: A Proxy War over foreign relations?}, url={https://melbourneasiareview.edu.au/covid-19-media-reporting-in-taiwan-a-proxy-war-over-foreign-relations/?print=pdf}, journal={Melbourne Asia Review}, publisher={Asia Institute}, author={Chang, Jasmine Li-Chia}, year={2020}, month={Aug}}

@article{cox1994seat,
  title={Seat bonuses under the single nontransferable vote system: Evidence from Japan and Taiwan},
  author={Cox, Gary W and Niou, Emerson},
  journal={Comparative Politics},
  pages={221--236},
  year={1994},
  publisher={JSTOR}
}

@book{cox1997making,
  title={Making votes count: strategic coordination in the world's electoral systems},
  author={Cox, Gary W},
  year={1997},
  publisher={Cambridge University Press}
}

@article{newman2023digital,
  title={Digital News Report 2023},
  author={Newman, Nic and Fletcher, Richard and Eddy, Kirsten and Robertson, Craig T and Nielsen, Rasmus Kleis},
  year={2023},
  publisher={RISJ: Reuters Institute for the Study of Journalism}
}

@article{rauchfleisch2022taiwan,
    title={Taiwan's public discourse about disinformation: The role of journalism, academia, and politics},
    author={Rauchfleisch, Adrian and Tseng, Tzu-Hsuan and Kao, Jo-Ju and Liu, Yi-Ting},
    journal={Journalism Practice},
    pages={1--21},
    year={2022},
    publisher={Taylor \& Francis}
}

@misc{TaiwaneseChineseIdentity2023, url={https://esc.nccu.edu.tw/PageDoc/Detail?fid=7800&amp;id=6961}, journal={Taiwanese / Chinese Identity}, publisher={Election Study Center, NCCU}, year={2023}, month={Jun}}

\section*{Supplementary Materials}

\begin{table}[!htb]
\begin{tabular}{|lllll|}
\hline
\multicolumn{5}{|c|}{Love React}                                                                                                                                                                                                                                                                                                                                                                                                                      \\ \hline
\multicolumn{1}{|l|}{}             & \multicolumn{1}{c|}{\begin{tabular}[c]{@{}c@{}}Ko \\ (R\textasciicircum{}2 = 0.387)\end{tabular}} & \multicolumn{1}{c|}{\begin{tabular}[c]{@{}c@{}}Hou \\ (R\textasciicircum{}2 = 0.597)\end{tabular}} & \multicolumn{1}{c|}{\begin{tabular}[c]{@{}c@{}}Lai \\ (R\textasciicircum{}2 = 0.521)\end{tabular}} & \multicolumn{1}{c|}{\begin{tabular}[c]{@{}c@{}}Gou \\ (R\textasciicircum{}2 = 0.455)\end{tabular}} \\ \hline
\multicolumn{1}{|l|}{china}    & \multicolumn{1}{l|}{}                                                                             & \multicolumn{1}{l|}{0.5536*}                                                                       & \multicolumn{1}{l|}{}                                                                              &                                                                                                    \\ \hline
\multicolumn{1}{|l|}{usa}      & \multicolumn{1}{l|}{0.8774**}                                                                     & \multicolumn{1}{l|}{0.6477*}                                                                       & \multicolumn{1}{l|}{}                                                                              & 0.9562**                                                                                           \\ \hline
\multicolumn{1}{|l|}{metoo}    & \multicolumn{1}{l|}{}                                                                             & \multicolumn{1}{l|}{}                                                                              & \multicolumn{1}{l|}{}                                                                              &                                                                                                    \\ \hline
\multicolumn{1}{|l|}{housing}  & \multicolumn{1}{l|}{1.025**}                                                                      & \multicolumn{1}{l|}{}                                                                              & \multicolumn{1}{l|}{}                                                                              &                                                                                                    \\ \hline
\multicolumn{1}{|l|}{energy}   & \multicolumn{1}{l|}{0.9364**}                                                                     & \multicolumn{1}{l|}{}                                                                              & \multicolumn{1}{l|}{}                                                                              &                                                                                                    \\ \hline
\multicolumn{1}{|l|}{tech}     & \multicolumn{1}{l|}{0.8973**}                                                                     & \multicolumn{1}{l|}{0.8129*}                                                                       & \multicolumn{1}{l|}{0.5975***}                                                                     &                                                                                                    \\ \hline
\multicolumn{1}{|l|}{ecofin}   & \multicolumn{1}{l|}{0.5456**}                                                                     & \multicolumn{1}{l|}{}                                                                              & \multicolumn{1}{l|}{0.6186***}                                                                     & 0.7903*                                                                                            \\ \hline
\multicolumn{1}{|l|}{in\_group}    & \multicolumn{1}{l|}{1.5864***}                                                                    & \multicolumn{1}{l|}{1.622***}                                                                      & \multicolumn{1}{l|}{1.4018***}                                                                     & 1.4619*                                                                                            \\ \hline
\multicolumn{1}{|l|}{out\_group}   & \multicolumn{1}{l|}{1.1861***}                                                                    & \multicolumn{1}{l|}{1.3636***}                                                                     & \multicolumn{1}{l|}{}                                                                              & 1.196***                                                                                           \\ \hline
\multicolumn{1}{|l|}{in\_and\_out} & \multicolumn{1}{l|}{-1.8332***}                                                                   & \multicolumn{1}{l|}{-1.711***}                                                                     & \multicolumn{1}{l|}{}                                                                              & -2.0372*                                                                                           \\ \hline
\end{tabular} \caption{Regression results against the love reaction. Aligns with the results for total interactions.} \label{tab:reg-love}
\end{table}

\begin{figure}[htbp] 
	\centering
		\includegraphics[width=1.0\columnwidth]{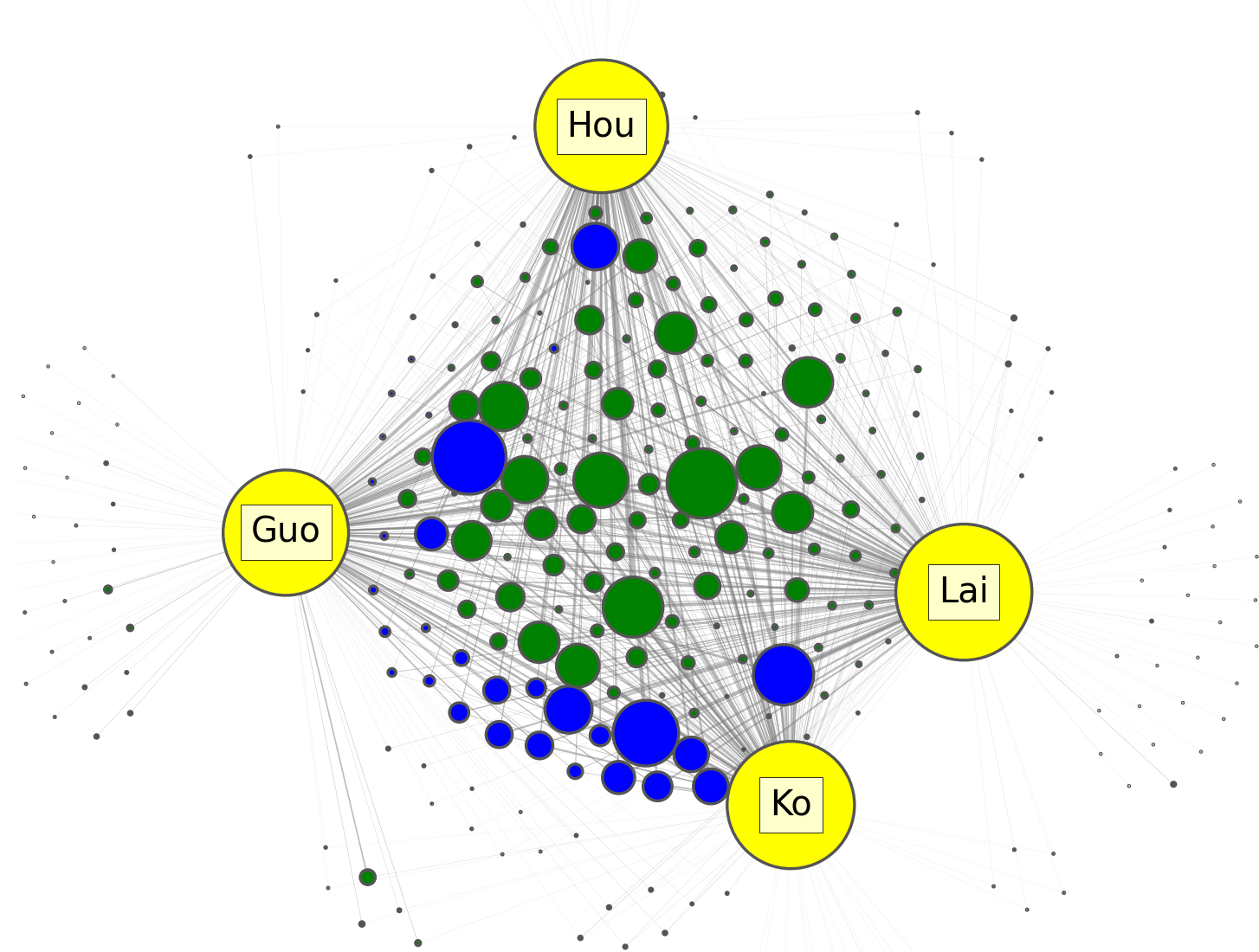}
	\caption{Network of presidential candidates (yellow), public groups containing Taiwan (green), and public groups containing R.O.C (blue).} \label{fig:simple_network}
\end{figure}

\begin{table}[!htb]
\begin{tabular}{|c|c|c|c|}
\hline
\multicolumn{4}{|c|}{Comparison of Keywords}              \\ \hline
Topic       & Candidate & Keyword List      & Love-to-Anger Ratio        \\ \hline
\multirow{4}{*}{USA}         & Lai       & \zh{主權、習近平、俄羅斯}        & 0.709 \\
\cline{2-4}            & Ko        & \zh{個案、本土、確診}          & 0.562 \\ 
\cline{2-4}            & Hou       & \zh{台獨、對話、交流、訪美、紐約、華府} & 0.773       \\ 
\cline{2-4}            & Gou       & \zh{經濟、發展、科技、利益}       & 0.722 \\\hline 
\multirow{4}{*}{Technology}  & Lai       & \zh{半導體、晶片、AI、台積電、戰爭}  & 0.781   \\
\cline{2-4}            & Ko        & \zh{城市、公園、公宅、台積電、基地}   & 0.700 \\ 
\cline{2-4}            & Hou       & \zh{和平、能源、環境、轉型、青年}    & 0.784 \\ 
\cline{2-4}            & Gou       & \zh{和平、能源、核電、鴻海、戰爭}    & 0.942 \\ \hline
\end{tabular} \caption{List of keywords by topic, with sentiment ratio.}
\end{table}

\end{document}